\documentclass[reprint,groupedaddress,superscriptaddress,longbibliography,tightenlines,nobibnotes,nofootinbib,aps,prb,floatfix]{revtex4-2}

\usepackage[utf8x]{inputenc}
\usepackage[english]{babel}
\usepackage[T1]{fontenc}
\usepackage{lmodern}
\usepackage{dsfont}
\usepackage{setspace}
\usepackage{parskip}
\usepackage[margin=0.6in]{geometry}
\usepackage{fancyhdr}
\usepackage{mathtools}
\usepackage{amsmath,amsfonts,amssymb,amsthm,bm,times}
\usepackage[breaklinks,colorlinks={true}, citecolor={blue}, filecolor={blue}, linkcolor={blue}, urlcolor={blue}]{hyperref}
\usepackage{url,graphicx,color}
\usepackage[caption=false]{subfig}
\usepackage{textcomp}
\usepackage{lipsum}
\usepackage{verbatim}
\usepackage{blindtext}
\usepackage{natbib}
\usepackage{makecell}

\usepackage{microtype}
\usepackage{braket}

\begin{document}

\title{Valley-controlled transport in graphene/ WSe$_{2}$ heterostructures under an off-resonant polarized light}
%%%%%%%%%%%%%%
\author{M. Zubair}
\email{muhammad.zubair@mail.concordia.ca; mzubair199@gmail.com}
\affiliation{Department of Physics, Concordia University, 7141 Sherbrooke Ouest, Montreal, Quebec H4B 1R6, Canada}
\author{P. Vasilopoulos }
\email{p.vasilopoulos@concordia.ca}
\affiliation{Department of Physics, Concordia University, 7141 Sherbrooke Ouest, Montreal, Quebec H4B 1R6, Canada}
\author{M. Tahir}
\email{tahir@colostate.edu; m.tahir06@alumni.imperial.ac.uk}
\affiliation{Department of Physics, Colorado State University, Fort Collins, CO 80523, USA}

\begin{abstract}

We investigate the electronic dispersion and transport properties of graphene/WSe$_{2}$ heterostructures in the presence of a proximity-induced spin-orbit coupling $\lambda_{v}$, sublattice potential $\Delta$, and an off-resonant circularly polarized light of frequency $\Omega$ that renormalizes  $\Delta$ to  $\bar{\Delta}_{\eta p} = \Delta +\eta p  \Delta_{\Omega} $ with $\eta$ %=\pm 1$ 
and $p$ %=\pm 1$} 
the valley and polarization indices, respectively, and $ \Delta_{\Omega} $  the gap due to the off-resonant circularly polarized light. Using a low-energy %effective 
Hamiltonian we find that the interplay between different perturbation terms leads to inverted spin-orbit coupled bands. At high $\Omega$ we study the band structure and dc transport using the Floquet theory and linear response formalism, respectively. We find that the inverted band structure transfers into the direct band one when the off-resonant light is present. The valley-Hall conductivity behaves as an even function of the Fermi energy in the presence and absence of this light. At $\Delta_{\Omega}$ = $\lambda_{v}$ - $\Delta$ a transition occurs from the valley-Hall phase to the anomalous Hall phase. In addition, the valley-Hall conductivity switches sign when the polarization of the off-resonant light changes. The valley polarization vanishes for $\Delta_{\Omega}$ = 0 but it is finite for $\Delta_{\Omega}$ $\neq$ 0 and reflects the lifting of the valley degeneracy of the energy levels, for $\Delta_{\Omega} \neq 0$, when the off-resonant light is present. The corresponding spin polarization, present for $\Delta_{\Omega}$ = 0, increases for $\Delta_{\Omega}$ $\neq$ 0. Further, pure $K$ or $K^{\prime}$ valley polarization is generated when $\Delta_{\Omega}$ changes sign.  Also, the %{\bf nondiagonal part of the} 
 charge Hall conductivity is finite for $\Delta_{\Omega}\neq 0$ and changes sign when  the handedness of the light polarization changes.

\end{abstract}

\maketitle

\section{introduction} \label{S1}

 Since its discovery graphene has attracted immense attention %of the scientific community 
 both theoretically and experimentally due to its peculiar electronic and optical properties~\cite{ii1}. But, it has limited usage in the field of spintronics due to its very weak intrinsic spin orbit coupling (SOC). The intrinsic SOC in graphene is  theoretically predicted to be weak, $12$ $\mu$eV~\cite{f7}. A value of $20$ $\mu$eV is reported in a recent experiment for graphene on SiO$_{2}$ substrate~\cite{ii2}. %In this way, 
 A lot of efforts have been made to enhance the strength of SOC in graphene by employing  external means, such as graphene hydrogenation~\cite{ii3,ii4} or fluorination~\cite{ii5} as well as heavy adatom decoration~\cite{ii6,ii7}, and bringing it to proximity with other two-dimensional materials specifically transition metal dichalcogenides (TMDCs)~\cite{ii8,ii9,ii10}. In recent years the heterostructures of graphene and TMDCs have become more promising because the Dirac cone of graphene is well fit in the band gap of TMDCs, which leaves it intact. The giant native SOC of TMDCs is transferred to graphene via hybridization processes. Moreover, the combinations of graphene with TMDCs, such as MoS$_{2}$ or WSe$_{2}$, exhibit the proximity SOC on the meV scale~\cite{nii0, nii1, nii2, nii3, nii4, nii5, nii6, f8}
% %nii7}
% , f8}.

Presently SOC, induced  by proximity effects,  is no longer limited to theoretical studies, as it has been demonstrated by experimentally as well~\cite{ii11}. The breaking of spatial symmetry due to the substrate leads to an alteration of the Hamiltonian and spin degeneracy of  graphene and opens a gap in its massless energy dispersion. In addition, it has been verified by experiments~\cite{ii12,f8,ii13,ii14} that another type of sublattice-resolved intrinsic SOC  arises, the so-called valley-Zeeman or staggered SOC with opposite sign on the $A$ and $B$ sublattices. Further, enhancement of the Rashba SOC and creation of staggered potentials are also unavoidable~\cite{f3}. 
 
Nowadays, the optical control of  functional materials has been become a hot topic in the condensed matter physics. In addition, it creates a bridge between  condensed matter physics~\cite{i1} and ultrafast spectroscopy~\cite{i2}. Many intriguing phenomena have been realized in  optically driven quantum solids such as light induced superconductivity~\cite{i3,i4}, 
photo-initiated insulator-metal transition~\cite{ 
   %i5,i6,
i7,i8}, microscopic interactions, such as the electron-phonon one, controlled by light~\cite{i9,i10,i11}, and theoretically predicted Floquet topological phases of matters~\cite{i12,i13,i14,i15,i16}. These Floquet  phases have  stimulated  much interest but   direct evidence for %the realization of 
electron-photon Floquet dressed states is scarce to date~\cite{i17,i18} contrary to the field of  artificial lattices~\cite{
% %i19, %i20,
i21, %i22,
i23,
%i24,%  %i25,i26,i27,
i28,i29, %i30,
%i31, 
% %i32,  %i33,
i34,i35}. 

Recently, light-induced anomalous Hall effect has been observed experimentally in monolayer graphene by using an ultrafast transport technique~\cite{i36} and predicted theoretically using a quantum Liouville equation with relaxation~\cite{i37}. Also, graphene under the influence of light has been studied in various frameworks~\cite{i12,i13,i14,i15,%i22,
i38,i39,i40,%i41,
i42, i44}
 %i43,i44}. 
The transport properties, especially valley-dependent dc transport, using the Floquet theory, has not been addressed sufficiently in contrast with a large amount of research on proximitized graphene. As far as transport in the presence of an off-resonant light is concerned, we are aware only of an electron transport study in MoS$_2$~\cite{f14}, of another one on graphene and the Lieb lattice~\cite{Tak}, and of a thermal transport study in  topological insulators in the absence of any SOC~\cite{tah}. Here we investigate theoretically the band structure in laser-driven graphene/WSe$_{2}$ heterostructures using the Floquet theory in the high-frequency regime. Also, we study dc transport in such heterostructres in the framework of linear response theory. We show that the interplay between the proximity SOCs and off-resonant light leads to a phase transition from the inverted band regime to the direct one. %Moreover, o
Our results are in good agreement with experimental results~\cite{i36} in the limit of vanishing proximity SOCs. 
    
In Sec.~\ref{S2} we specify the Hamiltonian and obtain the eigenvalues and eigenfunctions of the proximity modified graphene as well as an analytical expression for the density of states (DOS). In Sec.~\ref{S3} we derive analytical expressions for the conductivities and provide numerical results. Conclusions and a summary follow in Sec.~\ref{S4}.

%--------------------------------------------------------------------------------------------------------------------------------------------------------------------------------------------------
\section{Formulation}  \label{S2}

The real space tight-binding (TB) Hamiltonian of proximitized graphene is written as~\cite{f3, 
% %f2, 
f1, 
ff2}
%%%%%%%
\begin{eqnarray}
\notag
H & = & -t_J \sum_{\langle i,j \rangle,\alpha} c_{i\alpha}^{\dagger} c_{j\alpha} + \Delta \sum_{i \alpha}  \eta_{c_{i}} c_{i\alpha}^{\dagger} c_{i\alpha}  
\notag
\\ &&
+ \dfrac{i}{3 \sqrt{3}} \sum_{\langle \langle i,j \rangle\rangle,\alpha \alpha^{\prime}}  \lambda_{I}^{i} \nu_{ij} c_{i\alpha}^{\dagger} c_{j\alpha^{\prime}} [\bm{s_{z}}]_{\alpha \alpha^{\prime}}
\notag
\\ & &
+ \dfrac{2 i \lambda_{R}}{3} \sum_{\langle i,j \rangle,\alpha \alpha^{\prime}}  c_{i\alpha}^{\dagger} c_{j\alpha^{\prime}} [ (\bm{s} \times \mathbf{\hat{d}}_{ij})_{z}]_{\alpha \alpha^{\prime}}.
%\notag
%\\ & &
\label{e1}
\end{eqnarray} 
%%%%%%
Here $t_J$ is the hopping parameter, $c_{i\alpha}^{\dagger}$ creates an electron with spin polarization $\alpha$ at site $i$ that belongs to sublattice $A$ or $B$, and $\langle i,j \rangle$ $(\langle \langle i,j \rangle\rangle)$ runs over the nearest (second nearest) neighbouring sites. The second term is a staggered on-site potential, which takes into account the effective energy difference experienced by atoms at the lattice sites $A$ $(\eta_{c_{i}}=+1)$ and $B$ $(\eta_{c_{i}}=-1)$, respectively. The third and fourth terms represent the proximity-induced enhancement of the spin orbit coupling (SOC) due to a weak hybridization with the heavy atoms in TMDCs. The third term is the sublattice resolved intrinsic SOC ($\lambda_{I}^{i}$ with $i = A,B$) where $\nu_{ij}=+1$, if the second nearest hopping is anticlockwise, and $\nu_{ij}=-1$ if it is clockwise with respect to the positive $z$ axis. The  last term is the Rashba SOC parametrized by $\lambda_{R}$. It arises because the inversion symmetry is broken when the graphene sheet is placed on  top of TMDCs. Further,  $\bm{s}= (s_{x},s_{y},s_{z})$ is the Pauli spin matrix and $\mathbf{\hat{d}}_{ij}$ is the unit vector connecting the sites $i$ and $j$ in the same sublattice. 
%\linewidth \textheight
\begin{figure}[tp]
%\centering
%\hspace*{-0.4cm}
\centerline{\includegraphics[width=8cm, height=7cm]{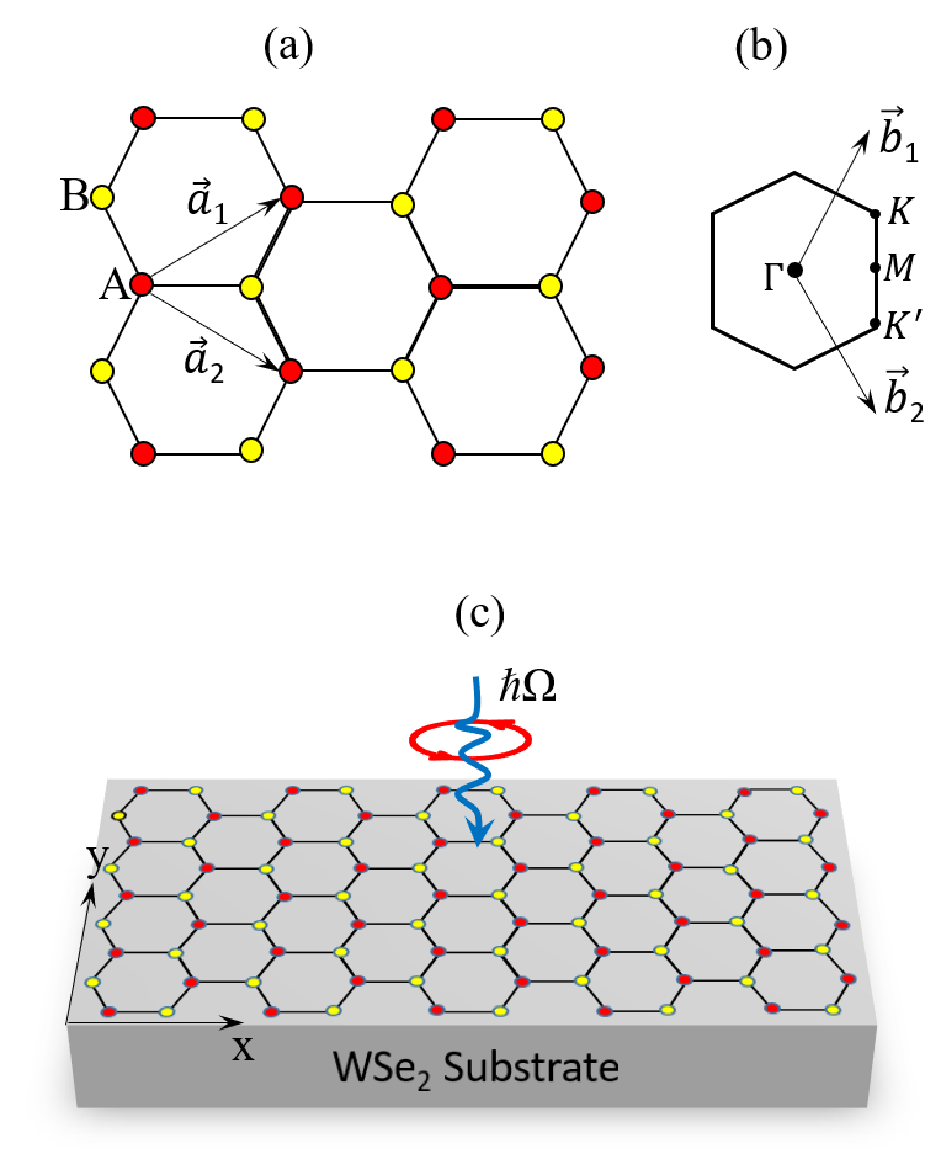}}
%\vspace{-0.2cm}
\caption{(a) Real-space graphene with $\vec{a}_{1}$ and $\vec{a}_{2}$ the primitive lattice vectors. (b) Graphene's first Brillouin zone and high symmetry points $\Gamma$, $K$, $K^{\prime}$,  and $M$  in reciprocal space. Its primitive lattice vectors are $\vec{b}_{1}$ and $\vec{b}_{2}$. (c) Schematics %representation 
of graphene epitaxially grown on a WSe$_{2}$ substrate and irradiated by a left circularly polarized light.}
\label{f01}
\end{figure}
%%%%%%%

We analyze the physics of electrons near the Fermi energy using a low-energy effective Hamiltonian derived from Eq.~(\ref{e1}) and a Dirac theory around $K$ and $K^{\prime}$ points. It  reads \cite{f4,f5,f6}
%%%%%%%%%%
\begin{eqnarray}
\notag
H_{{s_z} \eta} & = & v_{F}(\eta\sigma_{x}p_{x}+\sigma_{y}p_{y})+\Delta\sigma_{z}
+\lambda_{R}(\eta s_{y}\sigma_{x}-s_{x}\sigma_{y})
\\ & &
+ \dfrac{1}{2} [\lambda_{I}^{A}(\sigma_{z} + \sigma_{0}) + \lambda_{I}^{B}(\sigma_{z} - \sigma_{0})]\eta s_{z}
. \label{e2}
\end{eqnarray}
%%%%%%%%%

Here $\eta=+1(-1)$ denotes the valley $K$ ($K^{\prime}$), %and $K^{\prime}$, $s_{z}=+1(-1)$ is for spin up (down), 
 $\Delta$ is the mass term that breaks the inversion symmetry, $\lambda_{R}$  the Rashba type SOC strength, $\bm{\sigma}=(\sigma_{x}$, $\sigma_{y}$, $\sigma_{z})$ %($\sigma_{x}$, $\sigma_{y}$, and $\sigma_{z}$)  
 the Pauli matrix that corresponds to the pseudospin (i.e., $A-B$ sublattice); $\sigma_{0}$ is the unit matrix in the sublattice space and $v_{F}$ ($8.2 \times 10^{5}$ m/s) denotes the Fermi velocity of Dirac fermions. The last term arises due to the breaking of sublattice symmetry and can be categorized into two groups according to its dependence on sublattice spin: (i) $\lambda_{so} \sigma_{z} \eta s_{z} $ when $\lambda_{so}= (\lambda_{I}^{A}+\lambda_{I}^{B})/2$. This is called conventional Kane-Mele (KM) type SOC, which has a magnitude of the order of $\mu$eV in graphene/TMDCs heterostuctures \cite{f3,f6,f7}; (ii) $\lambda_{v} \sigma_{0} \eta s_{z} $ when $\lambda_{v}= (\lambda_{I}^{A}-\lambda_{I}^{B})/2$. It is called  valley-Zeeman or staggered SOC and has been experimentally confirmed in graphene on TMDCs \cite{ii12,f8,ii13,ii14};  it  occurs only for  $\lambda_{I}^{A}=-\lambda_{I}^{B}$. Further, Refs.~\cite{f7, f3, f6} show that 
$\lambda_{so}$ is negligibly small or zero. In view of that, we treat only the regime $\lambda_{v}>>\lambda_{so}$ and neglect $\lambda_{so}$ altogether. 
As shown %schematically
 in Fig.~\ref{f01}, monolayer graphene, irradiated by off-resonant circularly polarized light, is grown on WSe$_{2}$ that provides a staggered potential and induces SOC in graphene. We study the changes induced by circularly polarized light in graphene/WSe$_{2}$ in the presence of a perpendicular electric field $E$. We describe the monochromatic light through a  time-dependent vector potential $\vec{A}(t)= (E_{0}/\Omega)(\cos\Omega t, p\sin\Omega t)$ with $\Omega$ its frequency, $E_{0}$ the amplitude of the  field $E$, and $p = +1(-1)$ for left (right) circular polarization. The vector potential is periodic in time $A(t+T)=A(t)$ with $T=2\pi/ \Omega$. For high frequencies $\hbar\Omega \gg t_J$ and 
 low light intensities, %(ev_FE_0/\hbar\Omega)^2$,  
  i.e., ${\cal A}^{2} << 1$ with ${\cal A}=ev_FE_0/\hbar\Omega$ characterizing the intensity of light, Eq.~(\ref{e2})  gives the %effective static 
 Hamiltonian
 % Eq.~(\ref{e2}) in the presence of circularly polarized light reads
%%%%%%%%%%
\begin{eqnarray}
H_{s \eta}(t) & = &  H_{s \eta}^{0} + V(t) 
, \label{e3}
\end{eqnarray}
%%%%%%%%%
with
\begin{eqnarray}
\notag
H_{s_{z} \eta}^{0}  & = &   v_{F}(\eta\sigma_{x} p_{x}+\sigma_{y}p_{y}) 
 + \Delta  \sigma_{z} + \lambda_{v} \sigma_{0} \eta s_{z}
\notag
\\ & &
 +\lambda_{R}(\eta s_{y}\sigma_{x}-s_{x}\sigma_{y})
\notag
\\
V(t)  & = &  - %\dfrac{
(e v_{F}/\hslash)[\eta\sigma_{x} A_{x}(t)+\sigma_{y} A_{y}(t)]
. \label{e4}
\end{eqnarray}
%%%%%%%%%

 For $\hbar\Omega \gg t_J$ and ${\cal A}^{2} << 1$, %, where $t_J$ is the hopping parameter,
%For $\Omega >> t$ %the photon energy is much larger than the kinetic energy of the electrons
  Eq.~(\ref{e3}) can be reduced to an effective, %static, 
  time-independent Hamiltonian $H_{s_{z} \eta}^{\text{eff}}(t)$ using Floquet theory~\cite{i13}. $H_{s \eta}^{\text{eff}}(t)$ is defined through the time evolution operator over one period 
\begin{equation}
\hat{U} = \hat{T} \exp[-i \int_{0}^{T} H_{s_{z} \eta}(t) dt]= \exp[-i H_{s_{z} \eta}^{\text{eff}} T], \label{e5}
\end{equation}
%%%%%%%%%%
where $\hat{T}$ is time ordering operator. Using perturbation theory and expanding $\hat{U}$ in the limit of large frequency $\Omega$, we obtain
%%%%%
\begin{equation}
H_{s_{z} \eta}^{\text{eff}} = H_{s_{z} \eta}^{0} + %\dfrac{
[V_{-1} , V_{1}]/\hslash \Omega+ O(\Omega^{-2}), \label{e6}
\end{equation}
%%%%%
where $V_{m} = (1/T) \int_{0}^{T} e^{-im \Omega t} V(t) dt$ is the $m$-th Fourier harmonic of the time-periodic Hamiltonian and $[V_{-1} , V_{1}]$  the commutator between $V_{-1}$ and $V_{1}$. Corrections to Eq.~(\ref{e6}), to all orders of  $1/\Omega$, can be obtained by the method of Ref.~\cite{Tak}. Here we neglect them because we treat only the case $\hbar\Omega \gg t_J$. Using Eqs.~(\ref{e3}) and (\ref{e6}) we obtain
\begin{eqnarray}
\notag
H_{s_{z} \eta}^{\text{eff}} & = & v_{F}[\eta\sigma_{x} p_{x}+\sigma_{y}p_{y}] + \bar{\Delta}_{\eta p} \sigma_{z}+ \lambda_{v} \sigma_{0} \eta s_{z}
\\ & &
 +\lambda_{R}(\eta s_{y}\sigma_{x}-s_{x}\sigma_{y}),
 \label{e7}
\end{eqnarray}
%%%%%%%%%
%
\begin{figure}[tp]
%\centering
%\hspace*{-0.4cm}
\centerline{\includegraphics[width=\linewidth, height=\textheight,keepaspectratio]{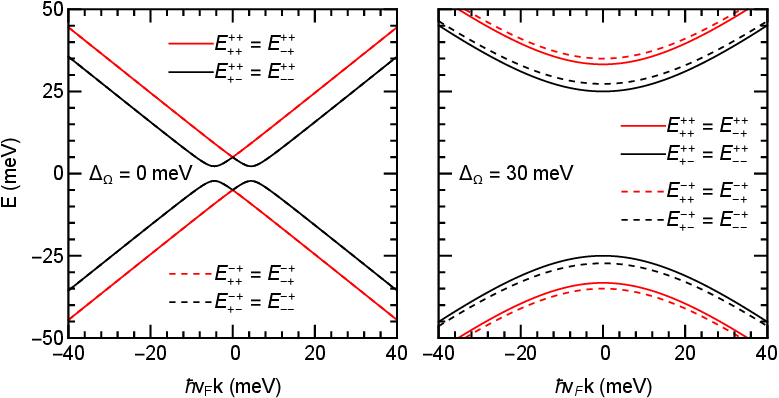}}
%\vspace{-0.2cm}
\caption{Energy dispersion curves around $K$ and $K^{\prime}$ of a graphene/WSe$_{2}$ heterostructure for $\Delta=1$ meV, $\lambda_{v}=4$ meV, and $\lambda_{R}=2$ meV. The left panel shows the inverted band regime, with strong spin mixing of different states%with black/red shading, 
, obtained for $\Delta_{\Omega}< \Delta + \lambda_{v}$. The right panel shows the direct band regime, with nearly full spin polarization, obtained %forn 
for $\Delta_{\Omega} > \Delta + \lambda_{v}$. The marking of all curves resulting from Eq.~(\ref{e8}), with  $p=1$ for all of them, %and not specified,
 is shown inside the panels. The solid black (red) curves are for $\eta=+1$ and $s=+1(-1)$ and the dashed black (red) ones  for $\eta=-1$ and $s=+1(-1)$.}
\label{f1}
\end{figure}
%%%%%%%
%where $\Delta_{\Omega}=v_{F}^{2}e^{2}E_{0}^{2}/\hslash \Omega^{3}$ is the %effective 
%energy term due to the circularly polarized light, which essentially renormalizes the mass of the Dirac Fermions~\cite{i13}. 
\hspace{-0.1cm} where $\bar{\Delta}_{\eta p}=\Delta + \eta p \Delta_{\Omega}$ with $\Delta_{\Omega}=v_{F}^{2}e^{2}E_{0}^{2}/\hslash \Omega^{3}$; $\bar{\Delta}_{\eta p}$ is the renormalized mass term  %$\Delta$ 
 due to the circularly polarized light which creates a gap $\Delta_{\Omega}$ in pure graphene, i.e., for $\Delta=0$, see  Ref. \cite{i13}.

The diagonalization of Eq.~(\ref{e7}) gives the dispersion 
%%%%%%%%%%%
\begin{eqnarray}
E_{\xi}^{\eta p}(k) & = & l \{G_{\eta} + 2 \lambda_{R}^{2}+ \epsilon_{k}^{2}  \label{e8}
%\\ & &
%\notag 
 + 2 s \sqrt{\Upsilon } \}^{1/2}.
\end{eqnarray}
where $\xi =\{l, s\}$ and $G_{\eta}= \lambda_{v}^{2} + \bar{\Delta}_{\eta p}^{2}$, $\Upsilon = \epsilon_{k}^{2} \bar{\lambda}^{2} + (\lambda_{R}^{2} - \lambda_{v}  \bar{\Delta}_{\eta p} )^{2} $ with $\epsilon_{k} = \hslash  v_{F} k$, $\bar{\Delta}_{\eta p}=\Delta + \eta p \Delta_{\Omega}$ and $\bar{\lambda}^{2} =  \lambda_{R}^{2} + \lambda_{v}^{2}$. Further, $l= +1 (-1)$ denotes the conduction (valence) band and $s= +1 (-1)$ represents the spin-up (spin-down) branches and is not a Pauli matrix $s_{z}$. The normalized eigenfunctions for both valleys are
%%%%%%%%%%%%
\begin{equation}
\psi_{\xi}^{+p} (k) = \dfrac{N_{\xi}^{+p}}{\sqrt{S_{0}}}
\begin{pmatrix}
1\\
A_{\xi}^{\eta p} e^{i\phi}\\
-i B_{\xi}^{\eta p} e^{i\phi}\\
-i C_{\xi}^{\eta p}  e^{2i\phi}
\end{pmatrix}
e^{i {\bf k} \cdot {\bf r}}\label{e9},
\end{equation}
%%%%%%%%%%%
\begin{equation}
\psi_{\xi}^{-p} (k) = \dfrac{N_{\xi}^{-p}}{\sqrt{S_{0}}}
\begin{pmatrix}
- A_{\xi}^{\eta p} e^{i\phi}\\
1\\
i C_{\xi}^{\eta p} e^{2i\phi}\\
-i B_{\xi}^{\eta p} e^{i\phi} 
\end{pmatrix}
e^{i {\bf k} \cdot {\bf r}}\label{e10},
\end{equation}
%%%%%%%%%%
respectively, with
\begin{align}
N_{\xi}^{\eta p} &  = l\big[1 + ( A_{\xi}^{\eta p}) ^{2} + ( B_{\xi}^{\eta p}) ^{2} + ( C_{\xi}^{\eta p}) ^{2}   \big ]^{-1/2},\label{e11}
\end{align}
 $S_{0}=L_{x}L_{y}$  the area of the sample, and $\phi = \tan^{-1}(k_{y}/k_{x})$. Further, 
$ A_{\xi}^{\eta p} =   \{ E_{\xi}^{\eta p} - \eta \alpha_{1}^{\eta} \}   /  \epsilon_{k}$, $B_{\xi}^{\eta p} = 2\lambda_{R}\{ (E_{\xi}^{\eta p})^{2} -  (\alpha_{1}^{\eta})^{2}  \} /  \epsilon_{k} \{  (  E_{\xi}^{\eta p} + \eta \alpha_{1}^{\eta})(  E_{\xi}^{\eta p} - \eta \alpha_{2}^{\eta}) - \epsilon_{k}^{2} \}$, and $C_{\xi}^{\eta p} =  2 \lambda_{R} \{ E_{\xi}^{\eta p} - \eta \alpha_{1}^{\eta} \} /  \{  (  E_{\xi}^{\eta p} + \eta \alpha_{1}^{\eta})(  E_{\xi}^{\eta p} - \eta \alpha_{2}^{\eta}) - \epsilon_{k}^{2} \} $ with $\alpha_{1}^{\eta}= \bar{\Delta}_{\eta p}+\lambda_{v}$, and $\alpha_{2}^{\eta}= \bar{\Delta}_{\eta p}-\lambda_{v}$.
%%%%%%
%
\begin{figure}[tp]
\centering
%\hspace*{-0.4cm}
\includegraphics[width=\linewidth, height=\textheight,keepaspectratio]{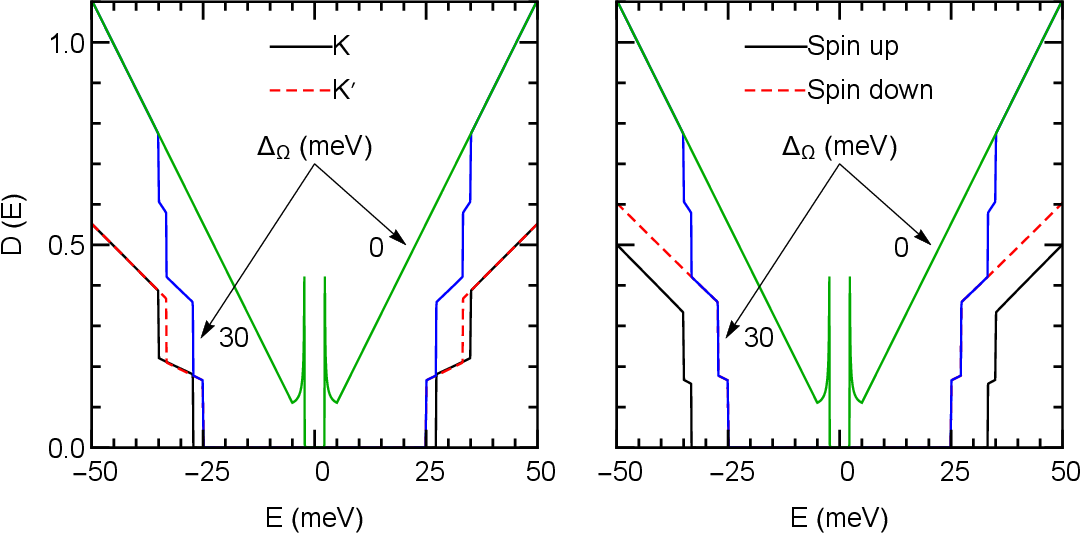}
\vspace{-0.4cm}
\caption{Density of states for two %different  
 values of $\Delta_{\Omega}$, as indicated, and $\Gamma=0.01$ meV. 
The left panel shows the valley components of the DOS, with both spins included, whereas the right panel shows the spin components of the DOS, with both valleys included. In both panels the  curves indicated by arrows show the total DOS. The  parameters 
$\Delta, \lambda_v $, and $\lambda_R $ are the same as in Fig.~\ref{f1}.
 The marking of the  curves  is shown inside the panels. In the left panel both spin contributions are included, in the right one both valley contributions are included.}
\label{f2}
\end{figure}
%%%%%%%

In numerical calculations throughout the manuscript, we use 
%we would use 
 values of the parameters $\Delta$, $\lambda_{v}$, and $\lambda_{R}$ somewhat larger than those of~\cite{f1} 
%{\color{red} The off-resonant light does not directly excite the electrons;  instead, it effectively modifies the electron bands through virtual photon absorption processes. To study the topological transitions of bands, this  light must satisfy the condition $\hslash \Omega \gg t$. Accordingly, in the  calculations we will use the values of $\Delta_{\Omega}$ from Refs. \onlinecite{i13,i36}.} 
%Here though, we will use somewhat larger values of these parameters as reported in~\cite{f1} 
to have well-resolved spin and valley splittings since the overall physics of the system is not changed when we do so. %use these slightly large values. 
  As for the  values of $\Delta_{\Omega}$,  it is known that the off-resonant light does not directly excite the electrons; instead, it  modifies the electron bands through virtual photon absorption processes. To study the topological transitions of bands, this light must satisfy the condition $\hslash \Omega \gg t_J$ and ${\cal A}^{2} << 1$. Accordingly, we will use the values of $\Delta_{\Omega}$ from Refs.~\onlinecite{i13,i36,f14}.%, and \onlinecite{f14}.

The typical band structure~(\ref{e8}) for both valleys is illustrated in Fig.~\ref{f1} for $p=+1$, $\Omega_{\Omega} < \Delta+\lambda_{v}$ (inverted band regime), and $\Delta_{\Omega} > \Delta+\lambda_{v}$ (direct band regime). The left panel shows the inverted band regime. The inversion occurs due to the anticrossing of the bands with opposite spins and in the presence of the Rashba SOC. The right panel depicts the direct band regime with simple parabolic dispersion. It is found that the spin and valley degeneracies are completely lifted when $\Delta_{\Omega}> \Delta + \lambda_{v}$, whereas the valley degeneracy is restored in the opposite limit similar to silicene~\cite{nc}. The valleys are interchanged %switch with each other 
if proximitized graphene is irradiated by a right circularly polarized light $p=-1$ (not shown here). 

\subsection{Limiting cases and density of states (DOS)}  \label{SS1}

i) Setting $\Delta=0$ in Eq.~(\ref{e8}), we obtain
%%%%%%%%%%%
\begin{eqnarray}
E_{\xi}^{\eta p}(k) & = & l \{ \lambda_{v}^{2} +\Delta_{\Omega}^{2}+ 2 \lambda_{R}^{2}+ \epsilon_{k}^{2}
%\\ & &
%\notag 
+ 2 s \sqrt{ Y } \}^{1/2}, \label{e12}
\end{eqnarray}
with $Y = \epsilon_{k}^{2}\bar{\lambda}^{2}+ (\lambda_{R}^{2} - \eta \lambda_{v} \Delta_{\Omega})^{2}$.

ii) In the limit $\lambda_{R}=0$, Eq.~(\ref{e8}) reduces
%%%%%%%%
\begin{equation}
E_{\xi}^{\eta p} (k) = l %\sqrt{
\big[\epsilon_{k}^{2} + \bar{\Delta}_{\eta p}^{2}\big]^{1/2} + s \lambda_{v}. \label{e13}
\end{equation}

The %density of states (
DOS per unit area corresponding to Eq.~(\ref{e8}) is given by 
%%%
\begin{eqnarray}
%\notag
\hspace*{-1cm}D(E)  =  \dfrac{\vert E \vert v_{F}^{-2}}{2 \pi \hslash^{2} } \sum_{\eta p}\Big[ \dfrac{ \theta(\vert E \vert - |E_{1g}^{\eta p}| )}{1 - \bar{\lambda}/ M^{+} }
+ \dfrac{ \theta(\vert E \vert -  |E_{2g}^{\eta p}| )} {1+ \bar{\lambda}/ M^{-} }
\Big],  \label{d1}
\end{eqnarray}
%%%%%%%%
with 
\begin{allowdisplaybreaks}
\begin{eqnarray}
\notag
\hspace*{-2cm}E_{1g}^{\eta p} & = & \lambda_{v}+ \bar{\Delta}_{\eta p}, \quad %\quad
%\notag
%\\ 
E_{2g}^{\eta p}  =  %\sqrt{ 
\big[(\lambda_{v}-\bar{\Delta}_{\eta p})^{2}+4\lambda_{R}^{2}\big]^{1/2}
\notag
\\ 
M^{\pm} & = &% \sqrt{
\big[(\lambda_{R}^{2}-\lambda_{v}\bar{\Delta}_{\eta p})^{2}+ \hslash ^2 v_{F}^{2} \bar{\lambda}^{2} \epsilon_{\pm}\big]^{1/2}
\label{d2}
%\notag
\\
%\hspace*{-1.4cm}
\hslash ^2 v_{F}^{2} \epsilon_{\pm} & = & E^{2}+\lambda_{v}^{2}-\bar{\Delta}_{\eta p}^{2} \pm 2
\notag
%\sqrt{
\big[\bar{\lambda}^{2} E^{2}-\lambda_{R}^{2}(\lambda_{v}+\bar{\Delta}_{\eta p})^{2}\big]^{1/2} .
\end{eqnarray}
\end{allowdisplaybreaks}

In Fig.~\ref{f2} we plot the DOS  given by Eq.~(\ref{d1}).
%using the Lorentzian form of the $\delta$ function $\pi \delta(x)= \lim_{\Gamma \to 0} \Gamma/(x^{2}+\Gamma^{2})$, where $\Gamma$ determines the amount of the broadening, in both %the invert and direct band regimes. 
The two jumps in the DOS indicate that two gaps open at each valley, displaying the clear signature of lifting the spin and valley degeneracies,  when graphene on WSe$_{2}$ substrate is in the direct band regime. The spin and valley degeneracies are completely lifted in the direct band regime while only the spin degeneracy is  lifted in the inverted band regime. Note that the DOS diverges in the inverted band regime as $D(E)\propto (E-\Delta_{1})^{-1/2}$ with $\Delta_{1}=\lambda_{R}(\lambda_{v}+ \Delta)/ (\lambda_{R}^{2}+\lambda_{v}^{2})^{1/2}$ (see green %, dot-dashed 
curves in both panels). This divergence is due to the Mexican-hat energy dispersion \cite{nc1}, cf. Fig.~\ref{f1}. 
In passing we may add that this behaviour of the DOS remains the same as the broadened one provided the level width $\Gamma$ is small, $\Gamma < 0.5$ meV. For higher $\Gamma$ the small structure of   the  DOS curves is smoothened out.
%%%%%%%
\begin{figure}[tp]
\centering
%\hspace*{-0.4cm}
\includegraphics[width=\linewidth, height=\textheight,keepaspectratio]{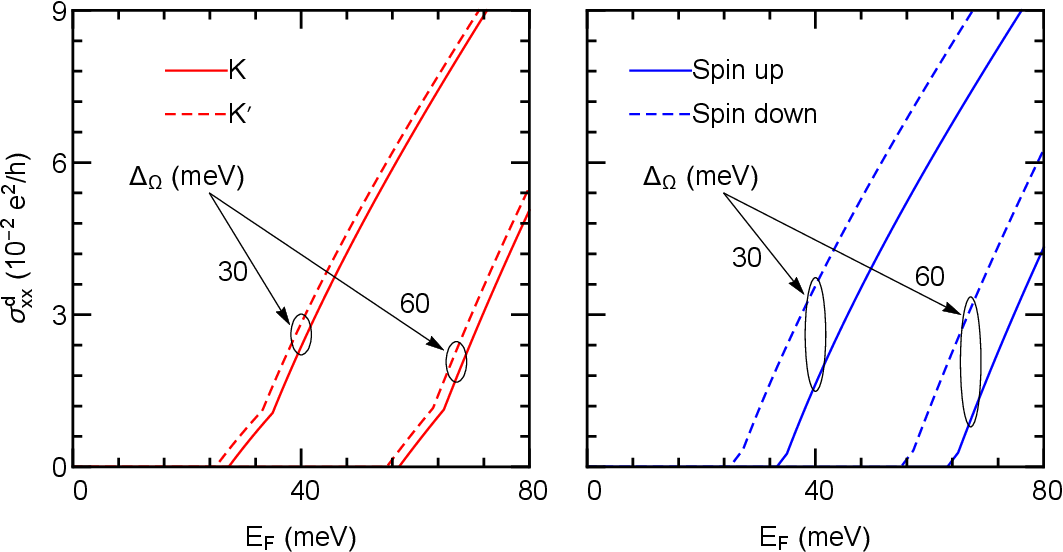}
%\vspace{-0.2cm}
\caption{Longitudinal conductivity vs Fermi energy $E_{F}$ for $T=0$ K, and $\tau_{F}=1 \times 10^{-15}$ sec. The other parameters are the same as in Fig.~\ref{f1}.}
\label{dia1}
\end{figure}
%%%%%%%

\section{ Conductivities}  \label{S3}

We consider a many-body system described by the Hamiltonian $H = H_{0} + H_{I} - \mathbf{R \cdot F (t)}$, where $H_{0}$ is the unperturbed part, $H_{I}=\lambda V$ is a binary-type interaction (e.g., between electrons and impurities or phonons) of strength $\lambda$, and $ \mathbf{- R \cdot F}(t)$ is the interaction of the system with the external field $F (t)$ \cite{f13}. For conductivity problems we have $\mathbf{F}(t) = e \mathbf{E}(t)$, where $\mathbf{E}(t)$ is the electric field, $e$ the electron charge, $\mathbf{R} = \sum_i {\bf r}_{i}$ , and $\mathbf{r}_{i}$  the position operator of electron $i$. In the representation in which $H_{0}$ is diagonal the many-body density operator $\rho = \rho^{d} + \rho^{nd}$ has a diagonal part $\rho^{d}$ and a nondiagonal part $\rho^{nd}$. Using $\rho = e^{-\beta H}$ and $H=H_0 +\lambda V$,    all operators were evaluated in the van Hove limit, $\lambda\to 0, t\to\infty$  but $\lambda^2 t$ finite, and all averages $<X>=Tr\{X\rho\}$ %were evaluated  
in the representation in which $H_0$ is diagonal. %Correspondingly, for weak ... , $\mu,\nu =x,y$".
 In this representation  $\lambda V$ is assumed nondiagonal; if it has a diagonal part, it's included in $H_0$.
Correspondingly, for weak electric fields and weak scattering potentials, for which the first Born approximation applies, the conductivity tensor has a diagonal part $\sigma_{\mu\nu}^{d}$ and a nondiagonal part $\sigma_{\mu\nu}^{nd}$; the total conductivity is $\sigma_{\mu\nu}^{tot} = \sigma_{\mu\nu}^{d} + \sigma_{\mu\nu}^{nd}, \mu,\nu = x,y$. For further details see Ref.  \cite{f13}.

 In general we have two kinds of currents, diffusive and hopping, with $\sigma_{\mu\nu}^{d} = \sigma_{\mu\nu}^{dif} + \sigma_{\mu\nu}^{col}$, but usually only one of them is present. The term  $\sigma_{\mu\nu}^{col}$ was introduced in Ref.~ \cite{f13} to distinguish collisional current contributions that are different from the standard diffusive ones valid for elastic scattering and characterized by a relaxation time $\tau$. As such, this is the main term for transport in a magnetic field when the diffusion contributions vanish. It also describes hopping between localized states. If no magnetic field is present, the hopping  term $\sigma_{\mu\nu}^{col}$ vanishes identically  and  only the term $ \sigma_{\mu\nu}^{dif} $ survives.  For  elastic scattering it is given  by~\cite{f13} 
%%%%%%%%
\begin{equation}
\sigma_{\mu\nu}^{d} = \dfrac{\beta e^{2}}{S_{0}} \sum_{\zeta} f_{\zeta} (1 - f_{\zeta} ) v_{\nu\zeta}\, v_{\mu\zeta}\, \tau_{\zeta} , \label{c1}
\end{equation}
%%%%%
with $\tau_{\zeta}$  the momentum relaxation time, and $v_{\mu\zeta}$ the diagonal matrix elements of the velocity operator. Further, $f_{\zeta} = \big[1 + \exp [\beta (E_{\zeta} - E_{F})]\big]^{-1}$ is the Fermi-Dirac distribution function, $\beta = 1/k_{B}T$, and $T$ the temperature.
%%%%% \linewidth

%%%%%%%
\begin{figure}[tp]
\centering
%\hspace*{-0.4cm}
\includegraphics[width=7.5cm, height=6.5cm,keepaspectratio]{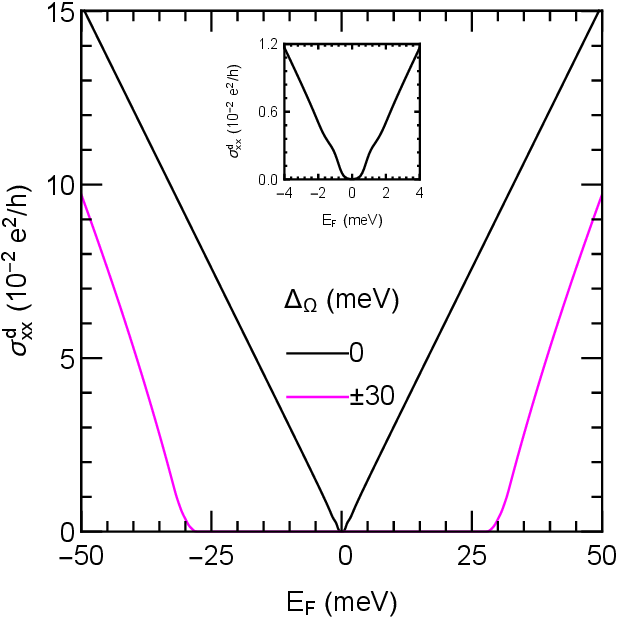}
%\vspace{-0.2cm}
\caption{Total longitudinal conductivity vs Fermi energy $E_{F}$. The parameters are $\Delta=0.54$ meV, $\lambda_{R} = 0.56$ meV, and $\lambda_{v} = 1.22$ meV \cite{f1}.}
\label{dia01}
\end{figure}
%%%%%%%

Regarding the contribution $\sigma_{\mu\nu}^{nd}$ one can use the identity $f_{\zeta} (1 - f_{\zeta^{\prime}})\big[1 - \exp [\beta (E_{\zeta} - E_{\zeta^{\prime}})]\big] = f_{\zeta} - f_{\zeta^{\prime}}$ and cast the original form \cite{f13} in the more familiar one 
%
%%%%
\begin{equation}
\hspace*{-0.4cm}\sigma_{\mu\nu}^{nd} = \dfrac{ i\hslash e^{2}}{S_{0}}\sum_{\zeta \neq \zeta^{\prime}} \dfrac{(f_{\zeta} - f_{\zeta^{\prime}}) \,v_{\nu\zeta\zeta^{\prime}} \,v_{\mu \zeta\zeta^{\prime}}}{(E_{\zeta} - E_{\zeta^{\prime}})(E_{\zeta} - E_{\zeta^{\prime}} - i \Gamma )} ,\label{c2}
\end{equation}
%%%%
where the sum runs over all quantum numbers $ \zeta $ and $ \zeta^{\prime} $ with $\zeta \neq \zeta^{\prime}$. The infinitesimal quantity $\epsilon$, in the original form of the conductivity,  has been replaced by $\Gamma_{\zeta}$ to 
%{\bf 
phenomenologically account for the broadening of the energy levels. 
%Note that we are considering the system in clean limit in the present study. 
One should keep in mind that a {\it strong} disorder may modify the Hall conductivity considerably. However, this problem is not studied here. In Eq.~(\ref{c2}) $v_{\nu \zeta \zeta^{\prime}}$ and $v_{\mu \zeta \zeta^{\prime}}$ are the off-diagonal matrix elements of the velocity operator. The relevant velocity operators are given by $v_{x}=  \partial H / \hslash \partial k_{x}$ and $v_{y}=  \partial H / \hslash \partial k_{y}$. With $\zeta=\{l, s, k, \eta, p \}=\{\xi, k, \eta, p \}$ for brevity, they read
%%%%%%
\begin{equation}
\left \langle \zeta  \right\vert  v_{x} \left \vert  \zeta^{\prime}\right\rangle  =  v_{F} N_{\xi}^{\eta p}N_{\xi^{\prime}}^{\eta p} (D_{\xi,\xi^{\prime}}^{\eta p} e^{i\phi} + F_{\xi,\xi^{\prime}}^{\eta p } e^{-i\phi} ) \delta_{\eta,\eta^{\prime}} \delta_{k,k^{\prime}}, \label{v1}
\end{equation}
%%%%%%
\begin{equation}
\left \langle \zeta^{\prime}  \right\vert  v_{y} \left \vert  \zeta \right\rangle  = i  v_{F} N_{\xi}^{\eta p}N_{\xi^{\prime}}^{\eta p} ( D_{\xi,\xi^{\prime}}^{\eta p} e^{-i\phi}  - F_{\xi,\xi^{\prime}}^{\eta p} e^{i\phi}  ) \delta_{\eta,\eta^{\prime}} \delta_{k,k^{\prime}}, \label{v2}
\end{equation}
where $D_{\xi,\xi^{\prime}}^{\eta p}= A_{\xi^{\prime}}^{\eta p}+ B_{\xi}^{\eta p}  C_{\xi^{\prime}}^{\eta p}$ and  $F_{\xi,\xi^{\prime}}^{\eta p}= A_{\xi}^{\eta p}+ B_{\xi^{\prime}}^{\eta p}  C_{\xi}^{\eta p}$.

The diagonal velocity matrix elements $v_{x\zeta}=\partial E_{\xi}^{\eta p}/\hslash \partial k_{x}$ from Eq. (\ref{e8}) can be readily found 
%%%%%%
\begin{eqnarray}
v_{x\zeta}= \dfrac{l \hslash v_{F}^{2}k_{x}}{E_{\xi}^{\eta p}} \big[ 1 + \dfrac{s\bar{\lambda}^{2}}{\sqrt{\Upsilon}}\big]. \label{v3}
\end{eqnarray}
%%%%%%
 
The above mentioned general expressions for conductivities are modified for Floquet theory \cite{i12} but are still valid for driven systems in the limit of large frequencies and weak intensity of light (${\cal A}<<1$) since only the zeroth level of the Floquet states contributes \cite{i13}, cf. Sec. III. Thus, these states can be taken as the eigenstates of Eq.~(\ref{e6}).  In addition, although Eq.~(\ref{e6}) is perturbative in $\Omega$, the above Hall conductivities expressions are nonperturbative in $\Omega$; that is, an infinitesimal gap $\bar{\Delta}_{\eta p}$ is sufficient to yield a topological band with a  quantized Hall conductance in unirs of $2 e^2/h$ \cite{i13}. Further the Fermi distribution is nonuniversal for systems which are out of equilibrium but for some cases of system-bath couplings \cite{flt2}, the steady-state distribution becomes thermal, and we restrict our results to such cases. Additionally, the electrode chemical potential will be small, for linear responses, compared to the intrinsic chemical potential of the system, and so we ignore the electrode chemical potential in our calculations. This allows us to write the chemical potential in the Kubo formalism as a constant, i.e. without accounting for sources at the boundaries. Also, it's worth pointing out that our approach for evaluating the conductivity tensor is the same or similar with that followed in Refs.~\cite{f14} for MoS$_2$, \cite{fr3, fr4} for silicene, and~\cite{ fr5} for WSe$_2$. In all of them a perpendicular electric field, not the source-to-drain one, was included in $H_0$. This is similar to our inclusion of the off-resonant light term $V(t)$ in $H_0$, as in the present work, and was also the case of Ref.~\cite{tah}.

%%%%%
\begin{figure}[tp]
\centering
%\hspace*{-0.4cm}
\includegraphics[width=\linewidth, height=7.5cm,keepaspectratio]{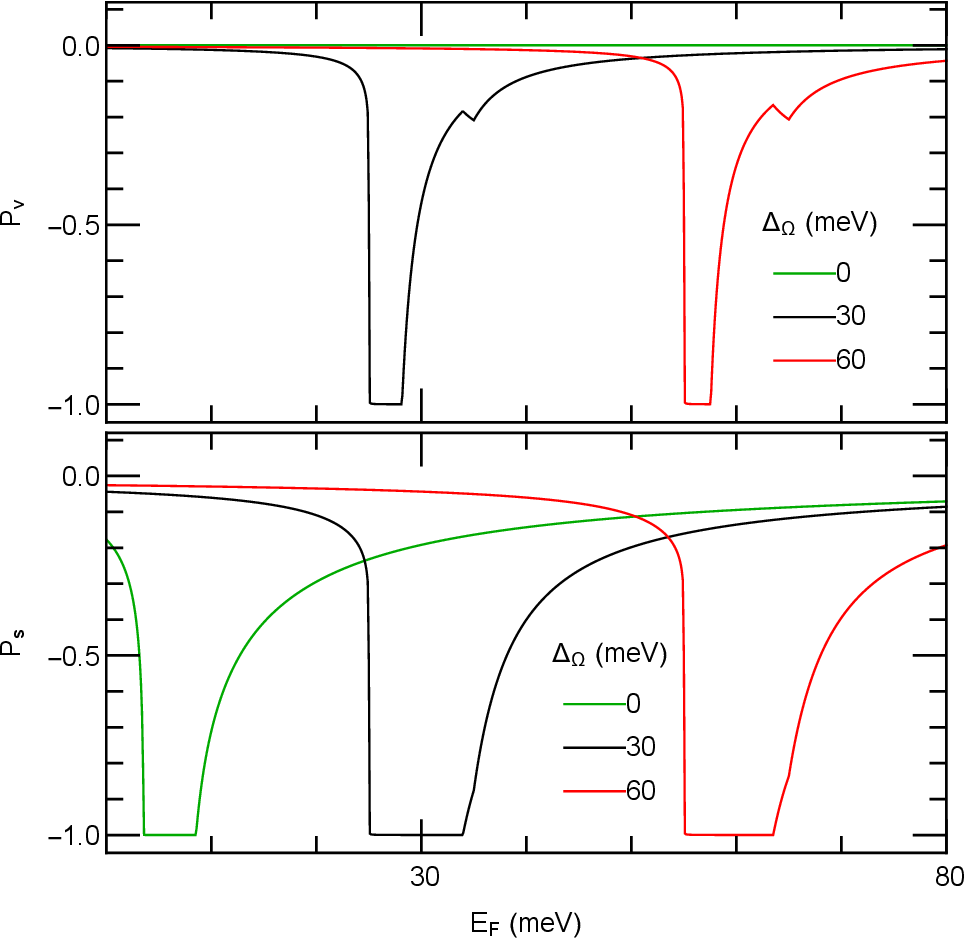}
\vspace{-0.2cm}
\caption{Valley $P_{v}$ and spin $P_{s}$ polarization vs $E_{F}$ for different values of $\Delta_{\Omega}$, as indicated, and  $\lambda_{R}=4$ meV. The other parameters are the same as in Fig.~\ref{dia1}.
 Notice that $P_{v}=0$ for $\Delta_{\Omega}=0$ while $P_{s}\neq 0$.}%height=6cm
\label{dia2}
\end{figure}
%%%%%%%

 We now calculate the conductivity $\sigma_{yx}^{nd}$ given by Eq.~(\ref{c2}). Further, the velocity matrix elements (\ref{v1}) and (\ref{v2}) are diagonal in $k$, therefore $k$ will be suppressed in order to simplify the notation. The summation in Eq.~(\ref{c2}) runs over all quantum numbers $\xi$, $\xi^{\prime}$, $\eta$, $\eta^{\prime}$, and $k$. The parameter $\Gamma_\zeta=\Gamma_{\eta \eta^{\prime}}^{\xi \xi^{\prime}}$, that takes into account the level broadening, is assumed  independent of the band and valley indices, i.e., $\Gamma_{\eta \eta^{\prime}}^{\xi \xi^{\prime}}=\Gamma$.  Using Eqs.~(\ref{v1}) and (\ref{v2}) we can express Eq.~(\ref{c2}) as
%%%%%%%%%
\begin{allowdisplaybreaks}
\begin{eqnarray}
\notag
\mathrm{Re}\sigma_{yx}^{nd}(\xi,\xi^{\prime},\eta, p) & = & \dfrac{2 e^{2} \hslash^{2} v_{F}^{2}}{h} %\sum_{\eta \xi \xi^{\prime}}
 \int dk k \,\dfrac{(N_{\xi}^{\eta p}N_{\xi^{\prime}}^{\eta p})^{2} (f_{\xi k}^{\eta p}- f_{\xi^{\prime}k}^{\eta p})}{  ( \Delta_{\xi \xi^{\prime}}^{\eta p})^{2}+ \Gamma^{2}}
\notag
\\ & &
\times  \big[(D_{\xi,\xi^{\prime}}^{\eta p})^{2} - (F_{\xi,\xi^{\prime}}^{\eta p})^{2}\big], 
\notag
\\
\mathrm{Im}\sigma_{yx}^{nd}(\xi,\xi^{\prime},\eta, p) & = & 0, \label{c5}
\end{eqnarray}
\end{allowdisplaybreaks}
%%%%%%
where $\Delta_{\xi \xi^{\prime}}^{\eta p}= E_{\xi k}^{\eta p} - E_{\xi^{\prime} k}^{\eta p}$. %The component $\sigma_{xx}^{nd}$ is also obtained from Eq. (\ref{c2}):

%Further, in the limit $\Gamma=0$, Eq. (\ref{c5}) reduces to
%%%%%%%%%%%%%
%\begin{eqnarray}
%\notag
%\mathrm{Re}\sigma_{yx}^{nd}(\xi,\xi^{\prime},\eta) & = & \dfrac{2e^{2} \hslash^{2} v_{F}^{2}}{h} %\sum_{\eta \xi \xi^{\prime}} 
%\int dk k\, \dfrac{(N_{\xi}^{\eta p}N_{\xi^{\prime}}^{\eta p})^{2} (f_{\xi k}^{\eta p}- f_{\xi^{\prime}k}^{\eta p})}{(\Delta_{\xi \xi^{\prime}}^{\eta})^{2}}
%\notag
%\\ & &
%\times \big[(D_{\xi,\xi^{\prime}}^{\eta })^{2} - (F_{\xi,\xi^{\prime}}^{\eta})^{2}\big]
%%\notag
%%\\
%%\mathrm{Re}\sigma_{xx}^{nd}(\xi,\xi^{\prime},\eta) & = & 0
%.\label{c3}
%\end{eqnarray}

 For $\lambda=\Delta= \Delta_{\Omega}=0$ and $\lambda_{R}\neq 0$, Eq.~(\ref{c5}) vanishes because the factor $(D_{\xi,\xi^{\prime}}^{\eta p})^{2} - (F_{\xi,\xi^{\prime}}^{\eta p})^{2}$ becomes zero. %whereas Eq. (\ref{c6}) survives. 
 Ignoring skew %scattering 
 and %other effects due to the 
 intervalley scatterings, the valley-Hall conductivity $(\sigma_{yx}^{v})$ %, spin-valley $(\sigma_{yx}^{sv})$, 
 %and spin $(\sigma_{yx}^{s})$ Hall conductivity 
 obtained from Eq.~(\ref{c5}) can be evaluated as
%%%%
\begin{equation}
\sigma_{yx}^{v} =  \sum_{\xi \xi^{\prime} p} \big[\sigma_{yx}^{nd}(\xi,\xi^{\prime}, +, p) 
 - \sigma_{yx}^{nd}(\xi,\xi^{\prime},-, p)  \big],\label{cc1}
%\\
%\sigma_{yx}^{s} & = & \sum_{\eta l \xi^{\prime}} \big[\sigma_{yx}^{nd}(l,+,\xi^{\prime},\eta) - \sigma_{yx}^{nd}(l,-,\xi^{\prime},\eta) \big],\label{cc2}
\end{equation}
%%%% 
where we set $\text{Re}\sigma_{yx}^{nd}(\xi,\xi^{\prime}, \eta, p ) \equiv \sigma_{yx}^{nd}(\xi,\xi^{\prime},\eta, p)$. The spin-Hall conductivity $\sigma_{yx}^{s}$ corresponding to Eq.~(\ref{c5}) is finite only when both KM and staggered SOCs are present~\cite{fff1}. Therefore, $\sigma_{yx}^{s}$ vanishes even in the presence of Rashba SOC. Even if it does not in graphene on WSe$_2$, it is assumed negligible in the regime $\lambda_v >> \lambda_{so}$ that we treat and we neglect it altogether, see also Sec.~\ref{S2}, above Eq.~(\ref{e3}). As usual, %Since a spin current is defined by $ \mathbf{J}_{s} = ( \hslash / 2e) (\mathbf{J}_{\uparrow} - \mathbf{J}_{\downarrow})$, 
we have to multiply $\sigma_{yx}^{v}$ by $1/2e$ ~\cite{ff2}. % , and $\sigma_{yx}^{s}$ by $\hslash/2e$
 %\cite{ff2}. % \cite{r5,r35}.

We can find a simple analytical result from Eq.~(\ref{cc1}) for the specific case $\lambda_{v}, \lambda_{R}=0$ in the low temperature limit. It is
%%%%%%%%%%
\begin{allowdisplaybreaks}
\begin{equation}
\sigma_{yx}^{v} = \begin{cases}
                        \dfrac{e}{2 h} , \  \quad -(\Delta + \eta p \Delta_{\Omega}) < E_{F} < \Delta + \eta p \Delta_{\Omega} \\ \\
                       \dfrac{e}{2 h} \dfrac{\eta \Delta + p \Delta_{\Omega}}{E_{F}},\   \quad E_{F} > \Delta + \eta p \Delta_{\Omega}
                 \end{cases} \label{cc5}
\end{equation}
\end{allowdisplaybreaks}
%%%%%%%%%%%

Eqs.~(16)-(17) of Ref.~\cite{f14} in the limit $\lambda \rightarrow 0$ are similar to  Eq.~(\ref{cc5}). %In the limit 
For $\Delta_{\Omega}\rightarrow 0$, Eq.~(\ref{cc5}) reduces to a result reported in Ref.~\cite{f15}. Further, we find the charge Hall conductivity 
%%%%%%%%%%
\begin{allowdisplaybreaks}
\begin{equation}
\sigma_{yx}^{c} = %\begin{cases}
                        %\displaystyle{
                        \sum_{p \eta \eta^{\prime} \xi \xi^{\prime}} \sigma_{yx}^{nd}(\xi,\xi^{\prime},\eta,\eta^{\prime}, p)= \begin{cases} 
                        0, &        \quad \Delta_{\Omega} = 0 \\ \\
                      % \displaystyle{ \sum_{\eta \eta^{\prime} \xi \xi^{\prime}}} \sigma_{yx}^{nd}(\xi,\xi^{\prime},\eta, \eta^{\prime})
                      \neq 0,              &  \quad \Delta_{\Omega} \neq 0
                 \end{cases} \label{cc6}
\end{equation}
\end{allowdisplaybreaks}
In the limit  $\Delta_{\Omega} \rightarrow 0$, $\sigma_{yx}^{c}$ vanishes. % and behaves like an odd function of $k$.  
%%%%%%
\begin{figure}[t]
\centering
\includegraphics[width=7.5cm, height=6.5cm,keepaspectratio]{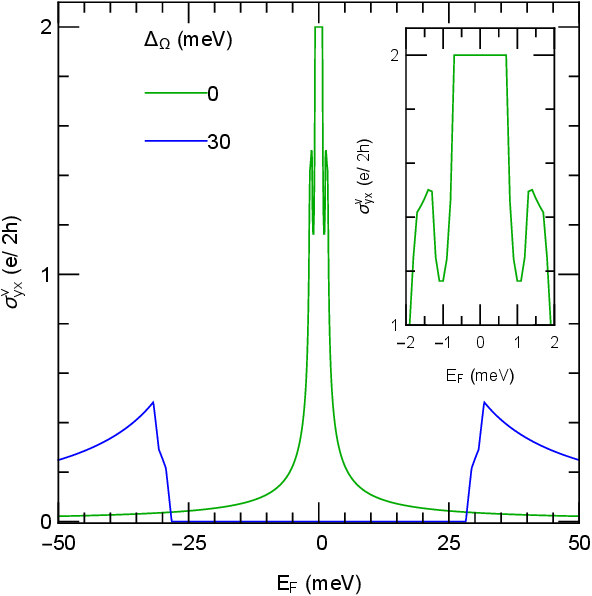}
\vspace{-0.2cm}
\caption{Valley-Hall conductivity vs. $E_{F}$ for $T=1$ K and $\Gamma=0$. The other parameters are $\Delta=0.54$ meV, $\lambda_{R} = 0.56$ meV, and $\lambda_{v} = 1.22$ meV~\cite{f1}. The green %dot-dashed 
 curve is measured in units of $e/h$  and the blue one in units of $e/10h$. The inset is a blowup of the region $-2$ meV $\leq E_F \leq 2$ meV. %of the green curve.
 }
\label{vh}
\end{figure}
%%%%%%%

We now consider the diagonal component $\sigma_{xx}^{d}$  given by Eq.~(\ref{c1}). Using Eq.~(\ref{v1}), with $\xi=\xi^{\prime}$, we obtain
%\begin{allowdisplaybreaks}
\begin{eqnarray}
\notag
\sigma_{xx}^{d}(\xi, \eta, p) & = & \dfrac{e^{2} v_{F}^{2} \beta}{\pi} %\sum_{\eta  \xi } 
\int dk k \,(N_{\xi}^{\eta p})^{4} f_{\xi k}^{\eta p} (1- f_{\xi k}^{\eta p}) 
\notag
\\ & &
\times (A_{ \xi}^{\eta p}+ B_{ \xi}^{\eta p}  C_{\xi}^{\eta p})^{2}\, \tau_{\xi k}^{\eta p}. \label{c4}
\end{eqnarray}
%\end{allowdisplaybreaks}

At very low temperatures we can make the approximation $\beta f_{\xi k}^{\eta p} (1- f_{\xi k}^{\eta p})\approx\delta(E_{\xi}^{\eta p}-E_{F})$ and  $\tau_{\xi k}^{\eta p}=\tau_{\xi k_{F}}^{\eta p}$. We find  $r=\sigma_{xx}^{nd}(\xi, \eta, p) / \sigma_{xx}^{d}(\xi, \eta, p) << 1$, mainly because %so we neglect 
 $\sigma_{xx}^{nd}(\xi, \eta, p) \propto \Gamma$. The precise value of $r$ depends on the scattering strength through $\Gamma$ and $\tau$ appearing in  $\sigma_{xx}^{d}(\xi, \eta, p)$. In what follows we neglect $\sigma_{xx}^{nd}(\xi, \eta, p)$.%For the $\tau$ used in Fig. 4, we have %indeed 
 %$r<<1$}. % because all states until the Fermi level are occupied. 
%\indent 

After evaluating the integral over $k$, Eq.~(\ref{c4}) becomes
%
%\begin{allowdisplaybreaks}
\begin{eqnarray}
\notag
%\hspace*{-0.3cm} 
\sigma_{xx}^{d}(\xi, \eta, p)  =  \dfrac{ e^{2} \tau_{F} E_{F}}{\pi \hslash^{2}%E_{F}^{-1}
}
 \Big[ 
% (A_{ \xi}^{\eta p}%(\epsilon_{0F}^{+})
%+ B_{ \xi}^{\eta p}%(\epsilon_{0F}^{+}) 
% C_{\xi}^{\eta p}%(\epsilon_{0F}^{+})
% )^{2} (N_{\xi}^{\eta p}%(\epsilon_{0F}^{+})
% )^{4} 
& Q_{ \xi}^{\eta p} &
  \dfrac{%(N_{\xi}^{\eta p}(\epsilon_{0F}^{+}))^{4} 
\theta( E_{F} - E_{1g}^{\eta p} )}{1 - \bar{\lambda}^{2}/ M }\Big |_{\epsilon_{+F}} 
\notag
\\ %\hspace*{-0.4cm} 
&+&%\hspace*{-0.4cm} &
%\times \dfrac{%(N_{\xi}^{\eta p}(\epsilon_{0F}^{+}))^{4} 
%\Theta( E_{F} -\vert E_{1g} \vert)}{1 - \bar{\lambda}^{2}/ M }|_{\epsilon_{0F}^{+}} 
%\notag
%\\ &&
% (A_{ \xi}^{\eta p}%(\epsilon_{0F}^{-})
%+ B_{ \xi}^{\eta p}%(k_{0F}^{-}) 
% C_{\xi}^{\eta p}%(\epsilon_{0F}^{-})
%)^{2}
%(N_{\xi}^{\eta p}%(\epsilon_{0F}^{-})
%)^{4} \,
Q_{ \xi}^{\eta p}  \dfrac{% (N_{\xi}^{\eta p}(\epsilon_{0F}^{-}))^{4} 
\theta(E_{F} -   E_{2g}^{\eta p} )} {1+\bar{\lambda}^{2}/ M }\Big |_{\epsilon_{-F}}
%\notag
%\\ &&
%\times \dfrac{% (N_{\xi}^{\eta p}(\epsilon_{0F}^{-}))^{4} 
%\Theta(E_{F} - \vert  E_{2g} \vert)} {1+\bar{\lambda}^{2}/ M }\big |_{\epsilon_{0F}^{-}}
\Big], %\hspace*{-0.5cm} 
\label{c7}
\end{eqnarray}
%\end{allowdisplaybreaks}
%%%%%%%
where $Q_{ \xi}^{\eta p}=(A_{ \xi}^{\eta p}+ B_{ \xi}^{\eta p} C_{\xi}^{\eta p} )^{2} (N_{\xi}^{\eta p} )^{4}$ and $\tau_{F} \equiv \tau_{\xi k_{F}}^{\eta p}$ is the relaxation time evaluated at the Fermi level. % and $k_{0F}^{\pm}\equiv \epsilon_{\pm}$. 
As indicated, the 1st and 2nd line in the square brackets are to be evaluated at $\epsilon_{+F}$ and $\epsilon_{-F}$,
respectively,  where $\epsilon_{\pm F}$ is obtained from Eq.~(\ref{d2}) for $E = E_F$. To evaluate Eq.~(\ref{c4}) numerically we used a Lorentzian broadening  of %the $\delta$ function 
$ \delta(E_{\xi}^{\eta p}-E_{F})$. %given after Eq.~(\ref{d2}). %. That is,
%%%%
%\begin{equation}
%\delta(x)=\dfrac{1}{\pi} \lim_{\Gamma \to 0} \dfrac{\Gamma}{x^2 + \Gamma^{2}}. \label{c
%\end{equation}
%%%%  

The valley $P_{v}$ and spin $P_{s}$ polarizations, corresponding to Eq.~(\ref{c4}), are 
%%%%%
\begin{equation}
P_{v}=\sum_{\xi p}\dfrac{\sigma_{xx}^{d}(l,s,+,p)-\sigma_{xx}^{d}(l,s,-,p)}{\sigma_{xx}^{d}(l,s,+,p)+\sigma_{xx}^{d}(l,s,-,p)}, \label{c9}
\end{equation}
%%%%% 
and
%%%%%
\begin{equation}
P_{s}=\sum_{\eta p l}\dfrac{\sigma_{xx}^{d}(l,+,\eta, p)-\sigma_{xx}^{d}(l,-,\eta, p)}{\sigma_{xx}^{d}(l,+,\eta, p)+\sigma_{xx}^{d}(l,-,\eta, p)}. \label{c10}
\end{equation}
%%%%%

In Fig.~\ref{dia1} we plot the conductivity, given by %illustrate the result of 
Eq.~(\ref{c4}), as a function of the Fermi energy $E_F$ by evaluating the integral over $k$ numerically for two values of the parameter $\Delta_{\Omega}$ and $p=+1$. Further, the left panel represents the valley-dependent contribution of Eq.~(\ref{c4}), with both spins included, whereas the right one depicts its spin-dependent contribution with both valleys included.
%Further, the left panel represents the valley part of Eq.~(\ref{c4}) with both spins included, whereas the right one depicts the spin part with both valleys included.
 To display the result clearly, we set $\Delta=1$ meV, $\lambda_{R}=2$ meV, $\lambda_{v}=4$ meV, and $\tau_{F}= 1 \times 10^{-15}$ sec. %in Fig.~\ref{dia1}. 
 We find that $\sigma_{xx}^{d}(\xi, \eta, p)$ vanishes when $E_F$ is in the gap while it increases linearly when $E_F$ is outside the gap. The kink appears when $E_F$ crosses the conduction band $(E_{++}^{\eta +})$. %Furtherm
 Moreover, we find $\sigma_{xx}^{d}(\xi, +, +) = \sigma_{xx}^{d}(\xi, -, +)$ in the inverted band regime $(\Delta_{\Omega}=0)$ while $\sigma_{xx}^{d}(\xi, +, +) \neq \sigma_{xx}^{d}(\xi, -, +)$ in the direct band regime $(\Delta_{\Omega} \neq 0)$. We %have 
 also verified that the analytical result (Eq.~(\ref{c7})) agrees well with the numerical one obtained from Eq.~(\ref{c4}).
%%%%%%%%%%%
\begin{figure}[t]
\centering
\includegraphics[width=7.75cm, height=5.75cm,keepaspectratio]{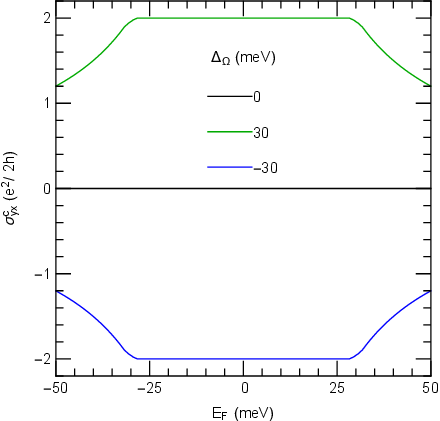}
\caption{Charge Hall conductivity vs. $E_{F}$ for different values of $\Delta_{\Omega}$. The other parameters are the same as in Fig.~\ref{vh}. It vanishes for $\Delta_{\Omega}=0$ and changes sign  when $\Delta_{\Omega}$ is changed to $ -\Delta_{\Omega}$.} %the charge Hall conductivity changes sign.}}
\label{vh1}
\end{figure}
%%%%%%%

%{\color{red} 
We plot the total longitudinal conductivity, with both valleys and spins included, in Fig.~\ref{dia01} for different values of $\Delta_{\Omega}$. As expected, %We found that 
$\sigma_{xx}^{d}$ is an even function of $\Delta_{\Omega}$. In addition, the band gap increases with %the increase of 
$\Delta_{\Omega}$.

The valley $P_{v}$ and spin $P_{s}$ polarizations versus $E_F$ are shown in Fig.~\ref{dia2} for $\lambda_{R}=4$ meV and three different values of %parameter 
$\Delta_{\Omega}$. It can be seen that $P_{v}= 0$ in the inverted band regime while $P_{v}\neq 0$ in the direct band one. In other words, the valley polarization can be switched on and off by controlling the parameter $\Delta_{\Omega}$. On the other hand, $P_{s} \neq %\geqslant 
0$ in both  band regimes. It is interesting to study $P_{v}$ in the direct band regime $(\Delta_{\Omega} \neq 0)$. The contribution of $\sigma_{xx}^{d}(\xi,+)$ to $P_{v}$ is zero in the range $\lambda_{v}+ \Delta - \Delta_{\Omega} \leqslant E_{F} < \lambda_{v}+ \Delta + \Delta_{\Omega}$. Thus, $P_{v}=1$, which is a pure $K^{\prime}$ valley polarization for $\Delta_{\Omega} \neq 0$. When we change the polarization of light to $p=-1$, a pure $K$ valley polarization is obtained. That is, one can easily reverse %generate   different types of 
the valley polarization by reversing that of the %direction 
circularly polarized light. This result may be useful in %the view point of 
valleytronics applications, such as making  valley valves~\cite{f16}.  

In Fig.~\ref{vh} we show the numerically evaluated valley-Hall conductivity $\sigma^v_{yx}$, from Eq.~(\ref{cc1}),
%the integral over wavevector $k$ 
in the inverted  $(\Delta_{\Omega} =0)$ and  direct  $(\Delta_{\Omega} \neq 0)$ band regimes  for %${\color{red} 
$l= l^{\prime}$ with $%\color{red}
s \neq s^{\prime}$, %{\it we also show it for} % 
as well as for %{\color{red} 
$l \neq l^{\prime}$ with $%{ \color{red} 
s = s^{\prime}$ and $%{\color{red}
s \neq s^{\prime} $. We used a sufficiently low temperature ($T=1$ K) to ensure that thermal vibrations of atoms have a negligible contribution to the electron transport. $\sigma^v_{yx}$ is quantized and has the universal value $2 e^{2}/h$ when the Fermi level is in the gap $-1$ meV $\leq E_F\leq 1$ meV (see green curve, compare with the DOS in Fig.~\ref{f1}). % (see green dot-dashed curve). 
 Its absolute value is reduced outside the gap as $E_{F}$ increases. The two peaks, to the left and right of the gap, at $E_F\approx \pm 1.5$ meV, appear due to the inverted band structure or the Mexican hat-like dispersion as can be seen in the inset of Fig.~\ref{vh}. $\sigma^v_{yx}$ vanishes 
when $E_{F}$ is in the gap in the direct band regime $\Delta_{\Omega} \neq 0$ as the blue curve shows. 
The reason is that in this case  electrons from both valleys flow in opposite directions and their contributions to the valley current exactly cancel each other. A non zero valley-Hall current is produced when $E_{F}$ crosses the conduction and valence bands. When $E_{F}$ grows further, the conductivity decreases. It is also worth noticing that the valley conductivity changes sign (not shown) if proximitized graphene is irradiated by a right circularly polarized light $(p=-1)$. 

For $\Delta_{\Omega} = 0$ a quantized valley-Hall conductivity of $2 e^{2}/h$ is obtained in the band gap as can be seen from the %dot-dashed 
 green curve in the inset of Fig.~\ref{vh}. On the other hand, for $\Delta_{\Omega} \neq 0$ the 
valley-Hall conductivity is quenched to zero within the band gap (see the blue curve of Fig.~\ref{vh}), while a quantized charge Hall conductivity of $2 e^{2}/h$ and $-2 e^{2}/h$ is obtained for the left- and right-handed circularly polarized light, respectively, as shown in Fig.~\ref{vh1}. The reason for the change $2 e^{2}/h\to -2 e^{2}/h$ is that 
this nondiagonal contribution to the conductivity is an odd function of $\Delta_{\Omega}$. 
\ \\
%factor $(N_{\xi}^{\eta p}N_{\xi^{\prime}}^{\eta p})^{2} [(D_{\xi,\xi^{\prime}}^{\eta})^{2} - (F_{\xi,\xi^{\prime}}^{\eta})^{2}]/( \Delta_{\xi \xi^{\prime}}^{\eta })^{2}$, called the Berry curvature, changes sign when $\Delta_{\Omega}$ is switched to $-\Delta_{\Omega}$.} It can be interpreted as a transition from a quantum valley Hall phase to a quantum anomalous Hall one for $\Delta_{\Omega} > \Delta + \lambda_{v}$. 

\section{Summary and conclusion}  \label{S4}

We %systematically 
investigated the valley-dependent dc transport by employing the linear response formalism and Floquet theory in the 
high-frequency limit as well as the energy dispersion in the presence of proximity-induced gaps. %under the influence of circularly polarized light. 
 We derived  analytical expressions for the energy dispersion relation of Dirac fermions, the DOS, and the diagonal and nondiagonal parts of the conductivity. We found that a transition occurs from an inverted band regime to a direct one for $\Delta_{\Omega} > \Delta + \lambda_{v}$ (see Fig.~\ref{f1}). In addition, the energy dispersion shows a complete  lifting of the {\it fourfold} spin and valley degeneracies in the direct band structure while it has a {\it twofold} valley degeneracy in the inverted band phase. We demonstrated that the DOS exhibits a van Hove singularity due to the inverted band structure, which   remained unchanged as long as $\Delta_{\Omega}< \Delta + \lambda_{v}$. The four jumps in the DOS are due to the lifting of the {\it fourfold} spin and valley degeneracy in the direct band regime in contrast to pristine graphene, cf.~Fig.~\ref{f2}. 

We showed that the valley polarization $P_{v}$ vanishes  for $\Delta_{\Omega} < \Delta + \lambda_{v}$ while for $\Delta_{\Omega} > \Delta + \lambda_{v}$ it is finite, $P_{v} \neq 0$;  this might be  useful in the design of   valleytronics devices such as optically controlled valley filters and valves based on  proxitimized graphene. On the other hand, $P_{s} \neq 0$ in both band regimes. 
%for both $\Delta_{\Omega} < \Delta + \lambda_{v}$ and $\Delta_{\Omega} > \Delta + \lambda_{v}$. 
Further, 100\% $K$ or $K^{\prime}$ valley polarization is achieved in the range $\lambda_{v}+ \Delta - \Delta_{\Omega} \leqslant E_{F} < \lambda_{v}+ \Delta + \Delta_{\Omega}$ when %we switch the sign of the circularly polarized light. 
 the handedness of the light polarization changes.

We found that, when $E_{F}$ in the gap, $\sigma_{yx}^{v}=2 e^{2}/h$ in the invert band regime while $\sigma_{yx}^{v}= 0$ in the direct band regime. Peaks are found in the curve of $\sigma_{yx}^{v}$ versus $E_F$ when $E_F$ crosses the inverted dispersion, see the %dot-dashed %dark
 green curve in Fig.~\ref{vh}. Moreover, for $\Delta_{\Omega}> \Delta + \lambda_{v}$, we have $\sigma_{yx}^{v} \neq 0$ when $E_F$ crosses the conduction and valence bands. The valley-Hall conductivity tends to $\sigma_{yx}^{v}= 0$ for both invert and direct band regimes in the limit $E_{F}\rightarrow \pm\infty$. A last finding is that the %{\bf nondiagonal part of the} 
 charge Hall conductivity is finite for $\Delta_{\Omega}\neq 0$ and changes sign when the handedness of the light polarization changes. 

Our results may be pertinent to developing future spintronics and valleytronics devices such as field-effect tunnelling transistors, memory devices, phototransistors, etc. 

\acknowledgments

M. Z. and P. V. acknowledge the support of the Concordia University Grant No.~NGR034 and a Concordia University Merit Fellowship. The work of M. T. was supported by Colorado State University.

\end{document}